\documentclass[5p,twocolumn]{elsarticle}

\usepackage{graphicx}
\usepackage{dcolumn}
\usepackage{pifont}
\usepackage{bm}
\usepackage{multirow}
\usepackage{amsmath}
\usepackage{float}
\usepackage{txfonts}
\usepackage[version=3]{mhchem}

\usepackage[usenames,dvipsnames]{xcolor}
\definecolor{blue}{RGB}{0,0,225}
\definecolor{cream}{RGB}{222,217,201}
\definecolor{red}{RGB}{225,0,0}

\journal{arXiv}

\begin{document}
\title{Contrary effect of B and N doping into graphene and graphene oxide heterostructures with \ce{MoS2} on interface function and hydrogen evolution}
	
\author[kimuniv-m]{Chol-Hyok Ri}
\author[kimuniv-m]{Yun-Sim Kim}
\author[kimuniv-m]{Kum-Chol Ri}
\author[kimuniv-m]{Un-Gi Jong}
\author[kimuniv-m]{Chol-Jun Yu\corref{cor}}
\cortext[cor]{Corresponding author}
\ead{cj.yu@ryongnamsan.edu.kp}

\address[kimuniv-m]{Chair of Computational Materials Design, Faculty of Materials Science, Kim Il Sung University, Ryongnam-Dong, Taesong District, Pyongyang, Democratic People's Republic of Korea}	

\begin{abstract}
Molybdenum disulfide (\ce{MoS2}) attracts attention as a high efficient and low cost photocatalyst for hydrogen production, but suffers from low conductance and high recombination rate of photo-generated charge carriers.
In this work, we investigate the \ce{MoS2} heterostructures with graphene variants (GVs), including graphene, graphene oxide, and their boron- and nitrogen-doped variants, by using first-principles calculations.
Systematic comparison between graphene and graphene oxide composites is performed, and contrary effect of B and N doping on interface function and hydrogen evolution is clarified.
We find that upon the formation of the interfaces some amount of electronic charge transfers from the GV side to the \ce{MoS2} layer, inducing the creation of interface dipole and the reduction of work function, which is more pronounced in the graphene oxide composites.
Moreover, our results reveal that N doping enhances the interface functions by forming donor-type interface states, whereas B doping reduces those functions by forming acceptor-type interface states.
However, the B-doped systems exhibit lower Gibbs free energy difference for hydrogen adsorption on GV side than the N-doped systems, which deserves much consideration in the design of new functional photocatalysts.
\end{abstract}

\maketitle

\section{Introduction}
Hydrogen production by using water and sunlight via photocatalysis draws considerable attention as a desirable solution to global environment and energy challenges.
In maximizing the efficiency of this solar-to-chemical energy conversion, photocatalysts are used to facilitate the hydrogen evolution reaction (HER)~\cite{Chen17nrm}.
While semiconductors are commonly adopted as such photocatalysts, generating electrons and holes with sunlight, they cannot give a high photoelectrochemical (PEC) activity without noble metal cocatalysts such as Pt, which retards the recombination of charge carriers.
However, the high cost and scarcity of noble metals limit their large-scale application.
During the past decade, therefore, the research attention has been focused on exploring noble-metal-free photocatalysts based on earth-abundant, low cost and nontoxic materials~\cite{Vesborg15jpcl}.

Among these, two dimensional (2D) layered transition metal dichalcogenides are recognized to be very promising candidates for HER photocatalysts~\cite{Sen20jpcl}.
In particular, \ce{MoS2} has attracted special interest due to its adequate band gap to visible light, high mobility of charge carriers and excellent light absorption property~\cite{Hu19jpcl,Li18an,Li18nl,Sun17nl}.
In fact, 2D \ce{MoS2} nanosheet has a direct band gap of 1.8 eV, which is more beneficial to visible light absorption than with the indirect smaller band gap of bulk phase (1.3 eV)~\cite{Song19jpcl,Mak10prl}.
With respect to the photocatalytic active sites of \ce{MoS2}, two different insights were established.
Firstly, the unsaturated S atoms at the exposed edge sites are argued to be the major active sites, thereby indicating that \ce{MoS2} monolayer with the richest edge sites can give the highest PEC activity~\cite{Wan18ne}.
Along with this, the basal plane S atoms were found to be active sites as well, emphasizing the electron hopping efficiency as the key for optimal PEC activity~\cite{Yu14nl}.
In this line, \ce{MoS2} heterostructures with graphene or reduced graphene oxide are expected to enhance the electron hopping efficiency and PEC activity by virtue of interfacial effects, such as suppressing charge recombination and improving charge transfer~\cite{Zou20jpcl,Liu20jpcl,Biroju17acs,Biroju17nt,Xu17cm,Li15sr,Chang14an,Zhou14aami}.
In design of the hybrid composites from \ce{MoS2} and graphene, there are two different ways according to which part induces HER.
Many works utilized \ce{MoS2} for a HER performer while graphene as a supplementary agent to overcome the poor electric conductivity of \ce{MoS2}~\cite{Li15sr,Chang14an,Liao14sr,Tang14sr}.
Contrary to this, the hydrogen production was designed to occur on the graphene side with the photo-generated electrons in the \ce{MoS2} side~\cite{Yang20jpcl,Biroju17acs,Biroju17nt}, by seeing that graphene variants (GVs) exhibit clear PEC activity~\cite{Albero19mole,Kumar18jmca,Li14jc,Yeh14am,Rao14nt}.
The detailed mechanism behind this reality needs more careful theoretical investigation.

\begin{figure*}[!th]
\centering
\includegraphics[clip=true,scale=0.09]{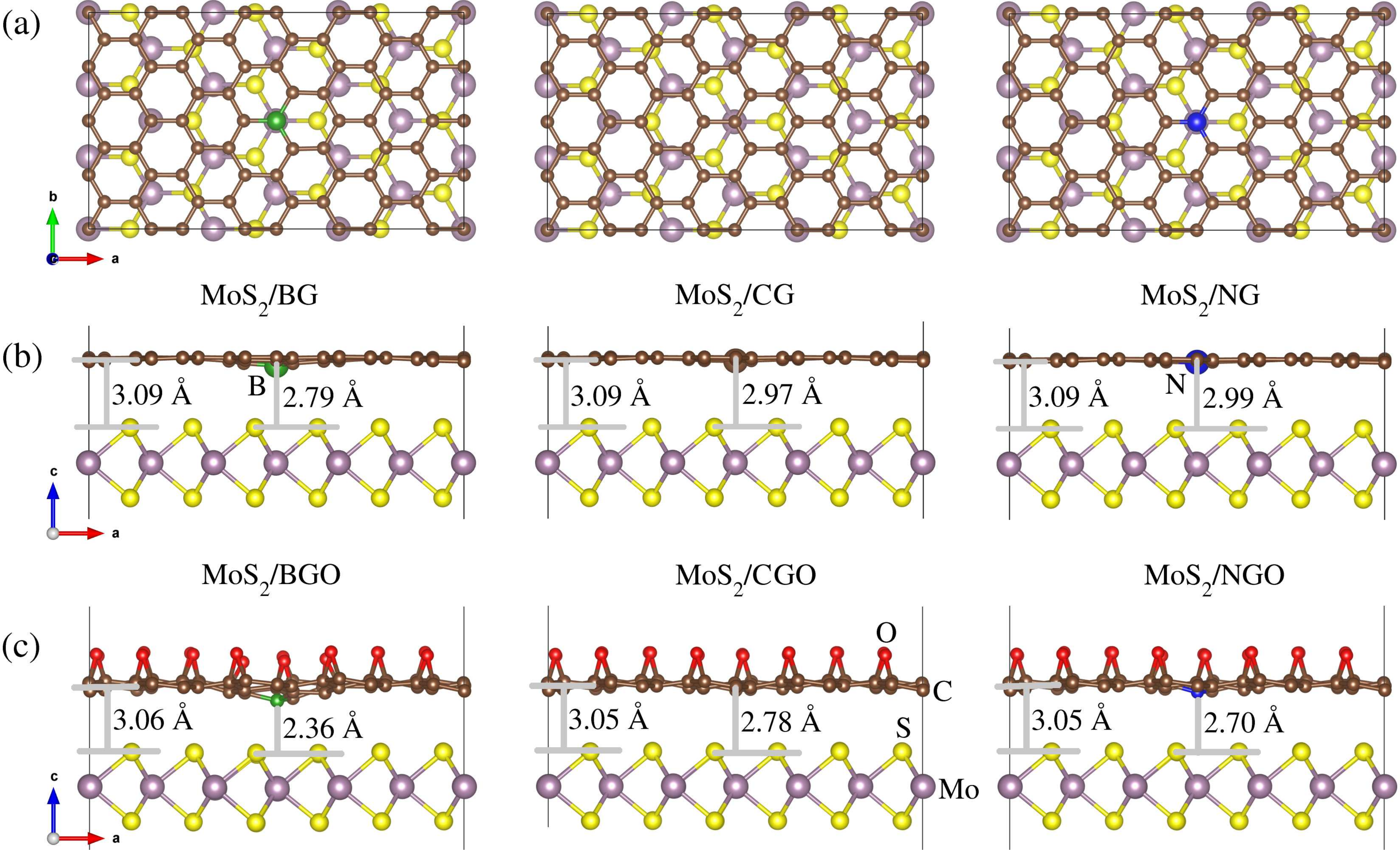} 
\includegraphics[clip=true,scale=0.45]{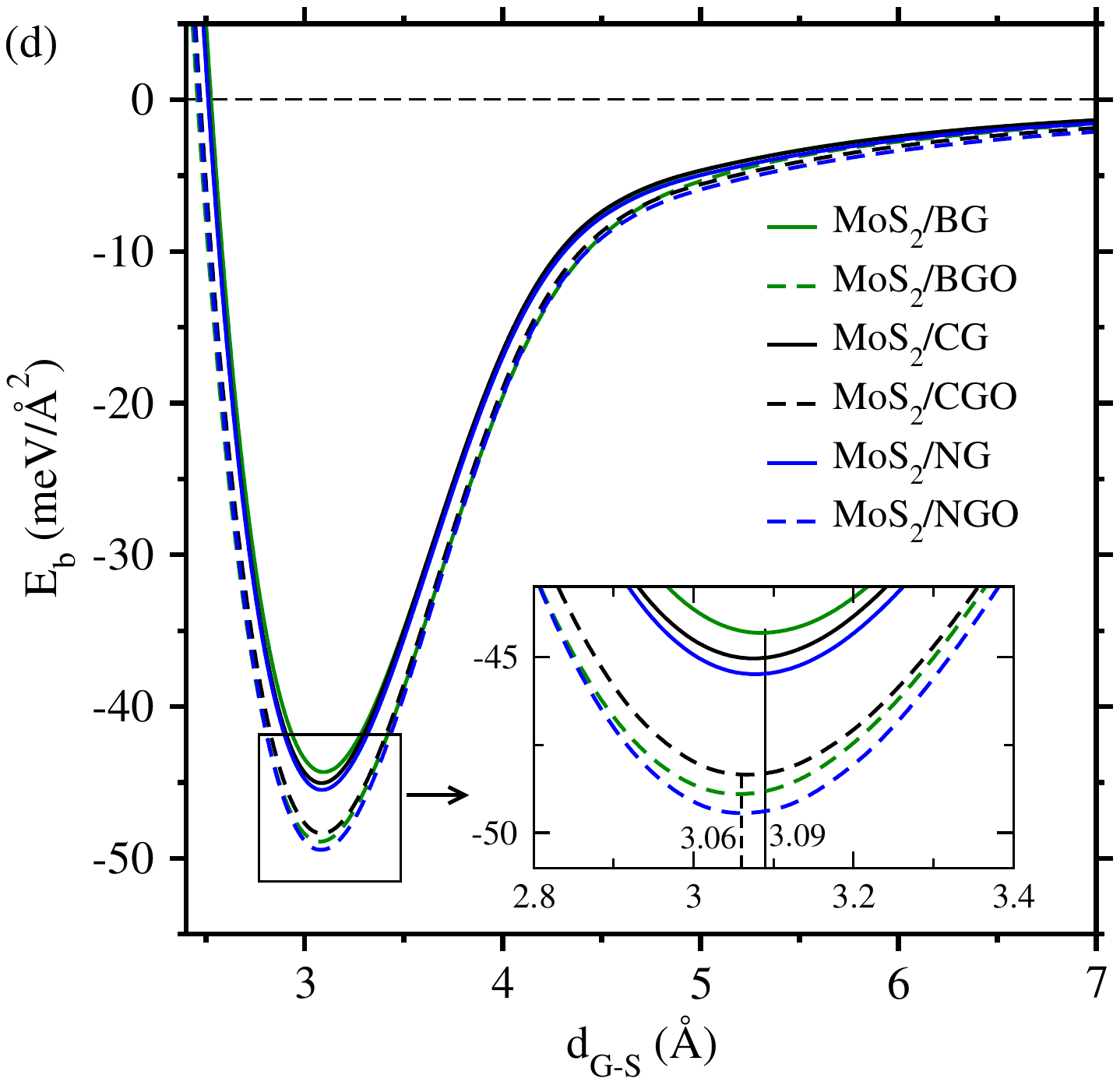}
\caption{Interface supercell geometries of \ce{MoS2} composites with B-doped graphene (BG), C-graphene (CG) and N-doped graphene (NG) in (a) top view and (b) side view, and with (c) B-doped graphene oxide (BGO), graphene oxide (CGO) and N-doped graphene oxide (NGO) in side view. 
The supercells are constructed of $3\times3$ \ce{MoS2} and $4\times4$ graphene or GO cells, and optimized by using the DZP basis sets and PBE + vdW functional.
The indicated distances are refereed to those from the average position of the topmost S atoms to the C atoms and the central C, B and N atom.
(d) Interlayer binding energy per carbon atom of the graphene or GO layer to the \ce{MoS2} layer, estimated by calculating the total energy difference of the interface system as increasing the interlayer distance from the system with infinity distance.
Inset shows the magnified view around the equilibrium distance.}
\label{fig1}
\end{figure*}

In this contribution, we demonstrated that the GV monolayer in the \ce{MoS2}/GV bilayer heterostructures can work as an efficient photocatalyst for \ce{H2} evolution with the electrons photo-generated and transferred from the \ce{MoS2} monolayer.
We tested graphene, reduced graphene oxide and their variants with $p$-type boron and $n$-type nitrogen doping, denoted as BG, CG, NG, BGO, CGO and NGO hereafter.
In fact, graphene itself cannot show photocatalytic activity due to its zero band gap, but such doping can give the finite band gap and higher electrical conductivity, resulting in the high PEC activity~\cite{Dong15sr,Tang16aem}.
Moreover, the functional epoxy groups ($-$O$-$) are expected to provide facile reaction sites for HER.
The defective doping, such as \ce{BCO2} and \ce{BC2O} for B doping or pyrrolic and pyridinic nitrogen for N doping, can play an important role, but they are not considered here according to our aim of revealing the effect of B or N doping on interface function and HER of \ce{MoS2}/GV hybrids.
We employed density functional theory (DFT) calculations considering van der Waals (vdW) interactions to study the \ce{MoS2}/GV bilayer heterostructures.
We presented quantitative change in binding energetics, electronic structure, interface function and hydrogen evolution by $p$-type B and $n$-type N doping.

\section{Methods}
Density functional theory calculations were carried out using the pseudopotential (PP) pseudoatomic orbital (PAO) basis set code SIESTA 4.1~\cite{SIESTA02jpcm}.
The double-$\zeta$ plus polarization (DZP) basis sets with the 300 meV energy shift for orbital-confining cutoff radii and 0.25 split norm for the split-valence were utilized for PAO basis set, as already applied to the interface~\cite{Ebnonnasir14apl}.
Using the ATOM code, we constructed the non-relativistic norm-conserving PPs with the valence electron configurations of H:$1s^1$, B:$2s^22p^1$, C:$2s^22p^2$, N:$2s^22p^3$ and O:$2s^22p^4$, while relativistic PPs for Mo:$5s^15p^04d^5$ and S:$3s^23p^4$.
The Perdew-Burke-Ernzerhof exchange-correlation functional~\cite{PBE}, together with the DFT-D2 vdW correction~\cite{Grimme06jcc}, was used to quantitatively treat both strong and weak binding interaction between layers and molecules.
In the present study, we considered a slab model of 2H polymorph of \ce{MoS2} sheet and GV layer.
The slab for \ce{MoS2} sheet has a $3\times 3$ periodicity in-plane and one (or two and three) \ce{MoS2} layer thick, while for the graphene sheet the slab consists of $4\times 4$ unit cells.
The vacuum layer of 50 \AA~thickness is added to the slab, which is large enough to prevent the artificial interaction between the periodic images.

The plane-wave cutoff energy for grid was set to 300 Ry and Monkhorst-Pack special $k$-points of $4\times6\times6$ and $6\times8\times1$ were used for bulk and slab models respectively.
For more accurate calculations of electronic density of states (DOS), we increased the $k$-points to $12\times16\times1$ for the interface models.
All atoms were relaxed until the atomic forces converged to 0.02 eV/\AA.
To eliminate the artificial electrostatic effects for asymmetric slabs, we took account of the slab dipole correction.
We computed the local DOS defined as the total DOS weighted with the electron density in the local region~\cite{Yu20acsami,Borlido18jctc}.
In the calculation of binding energy, we corrected basis set superposition error (BSSE) by applying the counterpoise method~\cite{Shuttleworth15jpcs}

\section{Results and discussion}
\subsection{Binding characteristics}
%structural property
We first explored the bonding nature of \ce{MoS2}/GV (GV = BG, CG, NG, BGO, CGO, NGO) bilayer heterostructures by calculating the geometric structures.
For all DFT calculations, we used the DFT-D2 vdW method~\cite{Grimme06jcc} in connection with the Perdew-Burke-Ernzerhof (PBE) functional~\cite{PBE} so as to treat the weak binding between the GV layer and the \ce{MoS2} layer.
The lattice constant and band gap of monolayer \ce{MoS2} with 2H polymorph were determined to be 3.19 \AA~and 1.7 eV (see Figure S2) in good agreement with the experimental and other DFT values of 3.20 \AA~and 1.8 eV~\cite{Song19jpcl,Mak10prl}.
It is worth noting that the band gaps of double and triple layer \ce{MoS2} were calculated to be 1.2 and 1.0 eV in direct mode, which are even lower than that of bulk (1.3 eV), thereby implying that the 2 or 3 layered \ce{MoS2} would have lower HER performance than the monolayer \ce{MoS2}~\cite{Biroju17acs}.
Also, the calculated lattice constants of CG and CGO sheet (2.46 and 2.51 \AA) are in good agreement with experiment.
The supercells were constructed using the $3\times3$ cell of \ce{MoS2} and $4\times4$ cell of CG or CGO with 4.6\% mismatch at most (see Figure S3 for their band structures and DOS).
A central carbon atom of CG and CGO was replaced with B or N atom to make the doping variant composite models.
Figures~\ref{fig1}a-c show the optimized structures.

\begin{table*}[!th]
%\centering
\caption{\label{tab1}Intralayer bond length $d_{\text{D}-\ce{C}}$ and interlayer distance $d_{\text{D}-\ce{S}}$ in \AA, interface binding energy $E_{\text{b}}$ and interlayer binding energy $E_{\text{int}}$ per area in meV/\AA$^2$, transferred electrons $\Delta q$ in e, interface dipole moment $\Delta\mu$ in Debye, and work function change $\Delta \phi$ in eV for \ce{MoS2} composites with graphene variants (GV).}
\small
\begin{tabular}{lccccccccccccc}
\hline
~~System & $d_{\ce{D}-\ce{C}}$ & $d_{\ce{D}-\ce{S}}$ &  $E_{\text{b}}$ & $E_{\text{int}}$ & $\Delta q_{\text{int}}$ & $\Delta q_{\ce{MoS2}}$ & $\Delta q_{\ce{GV}}$ & $\Delta\mu_{\text{int}}$ & $\Delta\mu_{\ce{MoS2}}$ & $\Delta\mu_{\ce{GV}}$ & $\Delta \phi_{\text{int}}$ & $\Delta\phi_{\ce{MoS2}}$ & $\Delta\phi_{\ce{GV}}$ \\
\hline
\ce{MoS2}/BG  & 1.47 & 2.79 & $-$30.66 & $-$44.26 & 0.0028 & 0.1674 & $-$0.1646 & 1.48 &$-$5.47 &~~6.95& 0.34 &$-$1.26 & 1.60 \\
\ce{MoS2}/CG  & 1.41 & 2.97 & $-$32.44 & $-$45.04 & 0.0037 & 0.2589 & $-$0.2552 & 2.25 &$-$7.97 &10.22 & 0.52 &$-$1.83 & 2.35 \\
\ce{MoS2}/NG  & 1.39 & 2.99 & $-$34.34 & $-$45.49 & 0.0044 & 0.3483 & $-$0.3439 & 3.04 &$-$10.39&13.43 & 0.70 &$-$2.39 & 3.09 \\
\ce{MoS2}/BGO & 1.50 & 2.36 & $-$19.77 & $-$48.89 & 0.0042 & 0.4162 & $-$0.4120 & 3.68 &$-$12.41&16.09 & 0.83 &$-$2.80 & 3.63 \\
\ce{MoS2}/CGO & 1.41 & 2.78 & $-$20.13 & $-$48.34 & 0.0055 & 0.5024 & $-$0.4969 & 4.20 &$-$14.93&19.13 & 0.95 &$-$3.37 & 4.32 \\
\ce{MoS2}/NGO & 1.40 & 2.70 & $-$22.87 & $-$49.43 & 0.0056 & 0.5331 & $-$0.5275 & 4.48 &$-$15.73&20.21 & 1.01 &$-$3.55 & 4.56 \\
\hline
\end{tabular} \\
$d_{\text{D}-\ce{C}}$ and $d_{\text{D}-\ce{S}}$ denote the bond length between the dopant atom and the nearest C atoms in the graphene or GO sheet and the distance between the dopant atom and the average position of topmost S atoms of the \ce{MoS2} layer. $E_{\text{b}}=\frac{1}{A}[E_{\ce{MoS2}/\ce{GV}}-(E_{\ce{MoS2}}+E_{\ce{GV}})+E_{\text{CP}}]$, where ``\ce{MoS2}/GV'', ``\ce{MoS2}'' and ``GV'' indicate the interface system, pristine \ce{MoS2} layer and graphene variant, respectively, and ``CP'' stands for counterpoise correction for BSSE.
$E_{\text{int}}= \frac{1}{A}[E(d_\text{eq})- E(d_\infty)]$, where $d_\text{eq}$ is the equilibrium interlayer distance and $d_\infty$ is approximated to be 30 \AA.
The transferred charge ($\Delta q=\int \Delta\bar{\rho}(z)dz$), interface dipole ($\Delta\mu=-\int z\Delta\bar{\rho}(z)dz$) and work function change ($\phi=e\mu/\varepsilon_0A$) are divided into two components: one from the \ce{MoS2} substrate and another from the GV layer.
The work function of the pristine \ce{MoS2} monolayer is calculated to be 4.04 eV, in good agreement with the experimental data (4.15 eV)~\cite{Kang19nml}.
\end{table*}

For the graphene composite, B and N doping were found to enlarge and contract the intralayer bond length between the dopant atom and its neighboring C atoms ($d_{\ce{D}-\ce{C}}$) from 1.41 \AA~to 1.47 and 1.39 \AA~respectively.
Meanwhile, they acted inversely for the interlayer distance between the dopant atom and the average position of the topmost S atoms ($d_{\ce{D}-\ce{S}}$) as from 2.97 to 2.79 and 2.99 \AA.
Another interlayer distance between the average position of C atoms and the topmost S atoms ($d_{\ce{G}-\ce{S}}$) was found to slightly increase from 3.086 \AA~for CG to 3.088 \AA~for BG and decrease to 3.084 \AA~for NG composite.
These indicate that B doping reduces the intralayer $\pi-\pi$ and interlayer vdW bondings, whereas N doping enhances these bondings.
For the cases of GO composites, similar change tendency for the intralayer bond length $d_{\ce{D}-\ce{C}}$ was observed as from 1.41 \AA~for CGO to 1.50 \AA~for BGO and 1.40 \AA~for NGO, but the interlayer bonding distance $d_{\ce{G}-\ce{S}}$ increased from 3.046 \AA~to 3.059 \AA~for both the B- and N-doped composites.

%binding energetics
To estimate the binding strength, we calculated the interface binding energy per area, $E_{\text{b}}=\frac{1}{A}(E_{\ce{MoS2}/\ce{G}}-E_{\ce{MoS2}}-E_{\ce{G}})$, and the interlayer binding energy per area, $E_{\text{int}}=\frac{1}{A}[E(d_\text{eq})-E(d_\infty)]$, where $A$ is the surface area and $d_\text{eq}$ is the equilibrium interlayer distance.
As listed in Table~\ref{tab1}, the interface binding energy in \ce{MoS2}/CG was calculated to be $-$35.23 meV/\AA$^2$, which is smaller than the previous DFT calculation of $-$51 meV/\AA$^2$~\cite{Ebnonnasir14apl}.
The interface binding energy of CGO composite ($-$22.83 meV/\AA$^2$) was smaller in magnitude than that of the CG composite.
For both the graphene and GO composites, B doping was found to reduce the interface binding, while N doping enhances the binding, in accordance with the change tendency of bond length.
As shown in Figure~\ref{fig1}d and listed in Table~\ref{tab1}, the strength of interlayer binding was in the order of BG $<$ CG $<$ NG for the graphene composites like the interface binding strength, but for the GO composites the order changed to CGO $<$ BGO $<$ NGO.
Such tendency of bonding strength change according to the central atom of GV sheet may be related with electron redistribution.

\begin{figure}[!th]
\centering
\includegraphics[clip=true,scale=0.115]{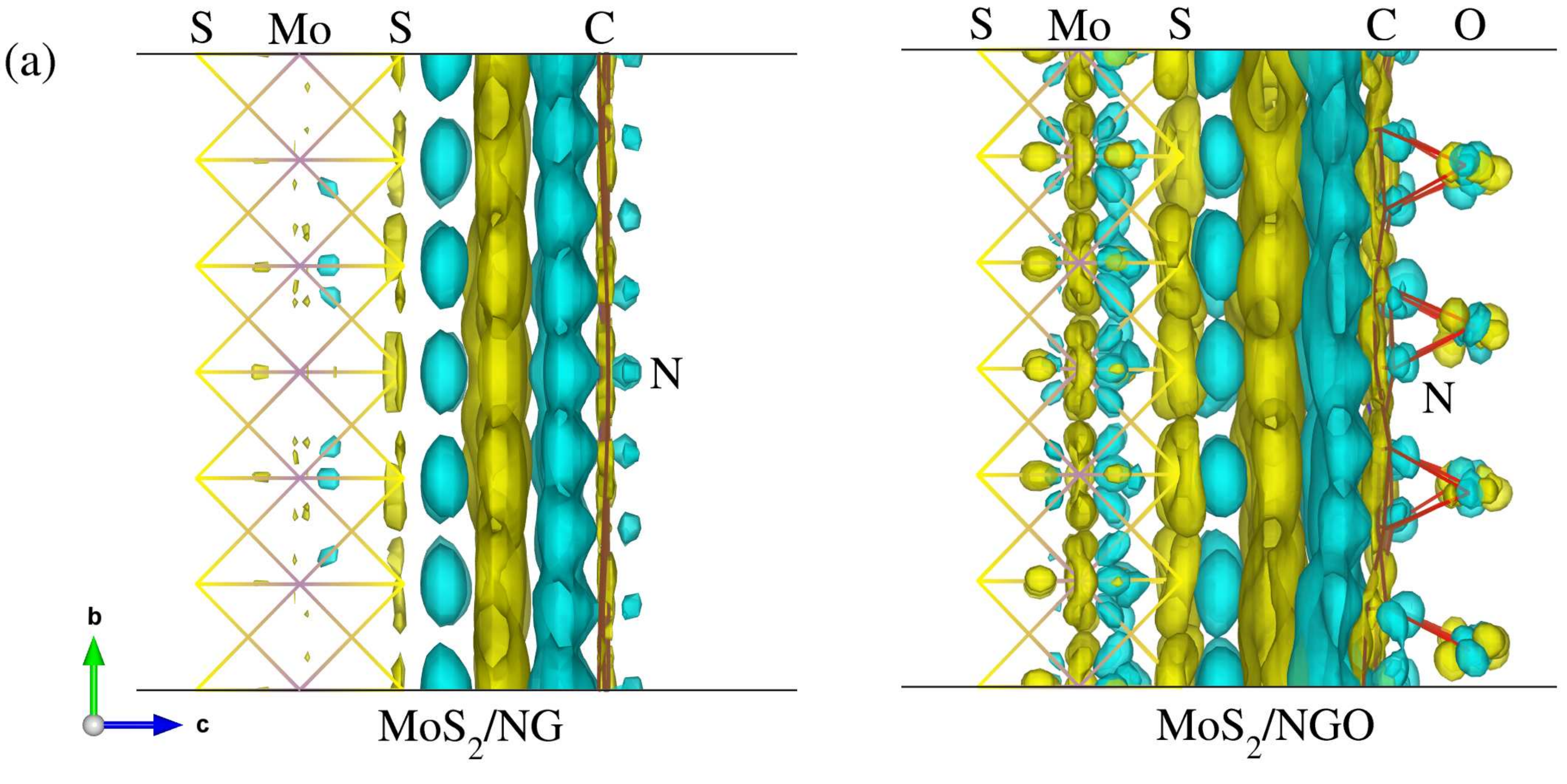} \\
\includegraphics[clip=true,scale=0.375]{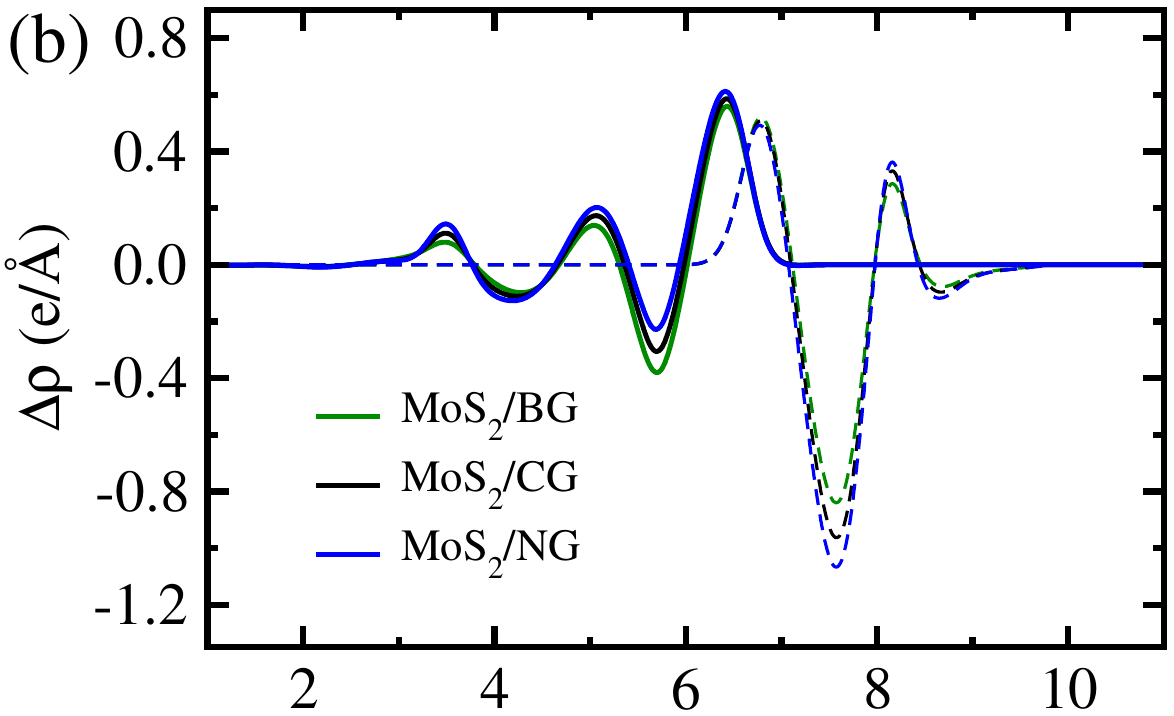}
\includegraphics[clip=true,scale=0.375]{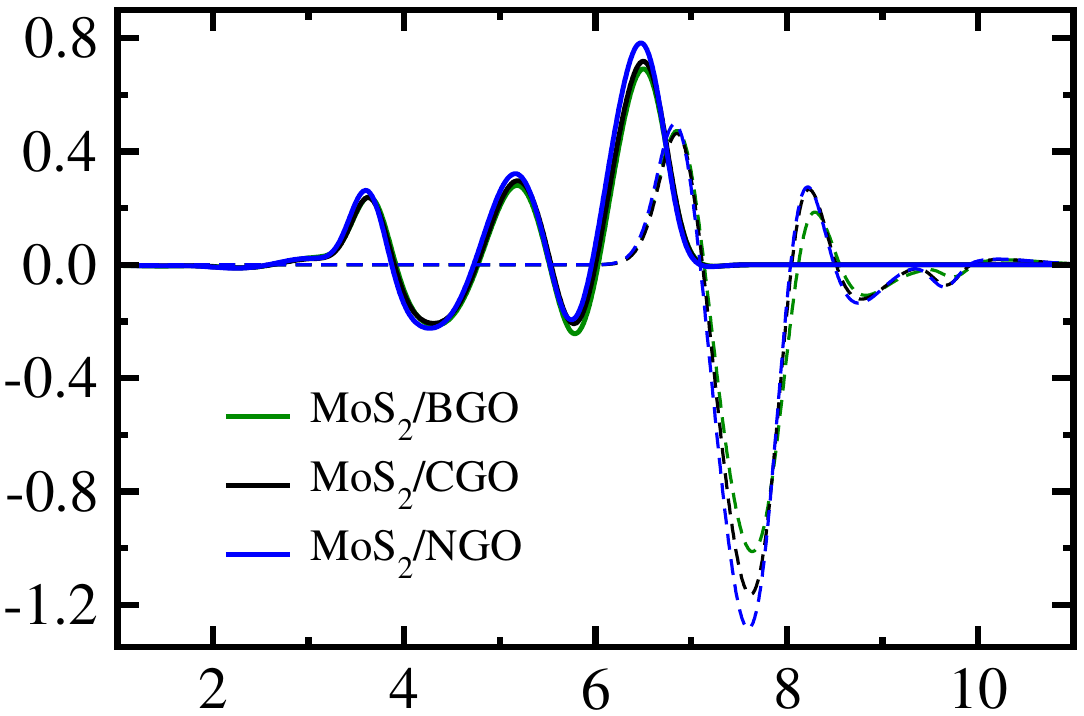} \\
\includegraphics[clip=true,scale=0.375]{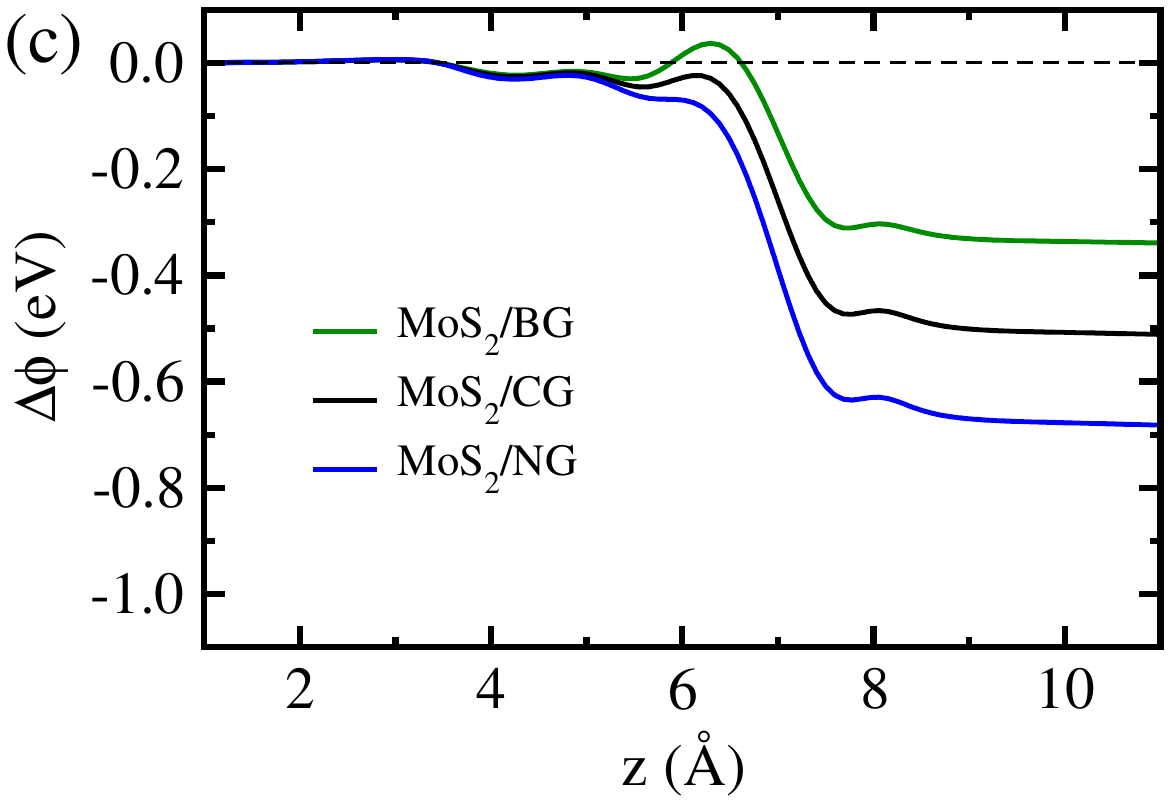}
\includegraphics[clip=true,scale=0.375]{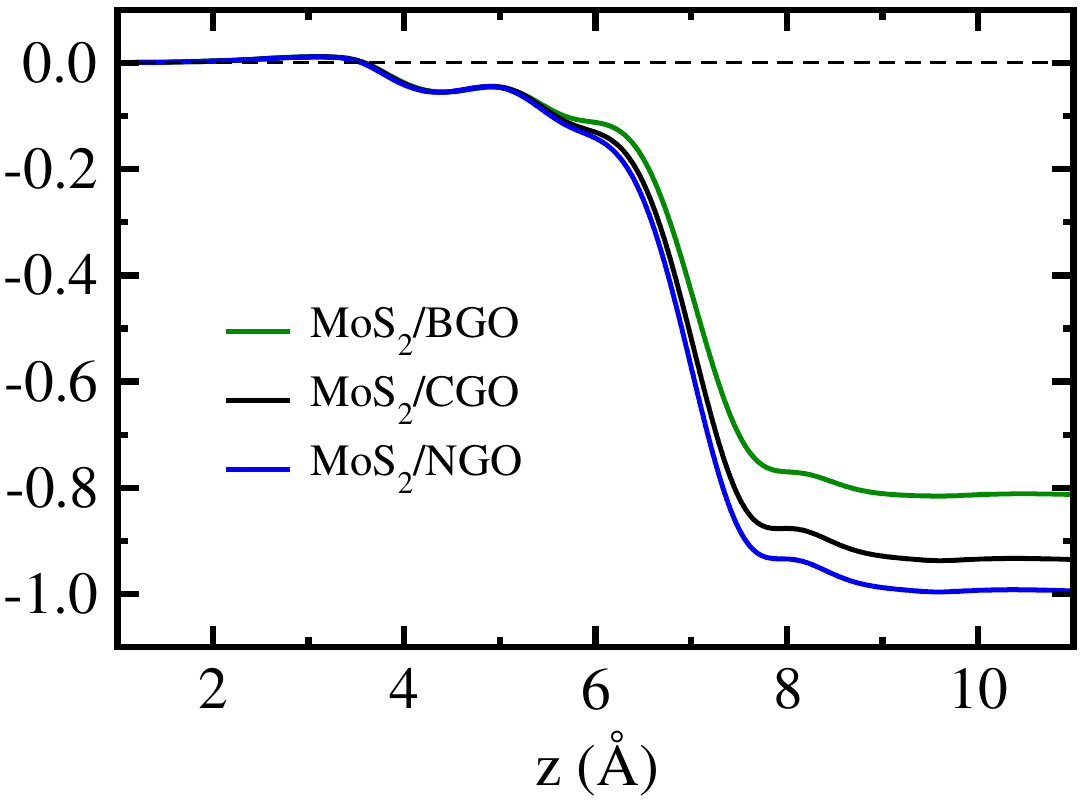}
\caption{(a) Isosurface view of spatial electron density difference at the value of 0.0006 e/\AA$^3$ upon the formation of \ce{MoS2}/NG and \ce{MoS2}/NGO interface systems, where yellow and cyan color indicate the electron accumulation and depletion, respectively.
(b) The $x$$-$$y$ plane-integrated electron density difference along the $z$-axis, which is divided into contributions from the \ce{MoS2} (solid line) and GV (dashed line) layers.
(c) Electrostatic potential change ($\Delta\phi$) estimated by solving the one-dimensional Poisson equation, $d^2\Delta\phi(z)/dz^2=\Delta\rho(z)$, with the plane-integrated electron density difference ($\Delta\rho$).}
\label{fig2}
\end{figure}

\subsection{Charge redistribution and interface dipole}
We show the spatial electron density difference plots at the same isosurface value of 0.0006 e/\AA$^3$ in Figure~\ref{fig2}a for \ce{MoS2}/NG and \ce{MoS2}/NGO selected among interface systems (see Figure S1 for the other interface systems).
Of the three different systems, the N-doped composites exhibit the most remarkable electron redistribution, while the B-doped systems show the least.
The electron accumulation and depletion regions, forming layers parallel to the interface, appear alternately, implying generation of interface dipole as discussed below.
To show more clearly, we depicted the $x$$-$$y$ plane-integrated electron density difference $\Delta\bar{\rho}(z)$ in Figure~\ref{fig2}b, being divided into contributions from the \ce{MoS2} and GV layers.
The significant electron accumulation, originated from both the \ce{MoS2} and GV layers, can be seen in the middle of interface region towards the \ce{MoS2} side, while the striking electron depletion is found in the GV side.
This indicates that some electrons are transferred from GV to \ce{MoS2}.
By integrating the electron density difference, $\Delta q=\int\Delta\bar{\rho}(z)dz$, we estimated the amount of transferred electrons $\Delta q$, as listed in Table~\ref{tab1}.
Confirming that \ce{MoS2} (GV) received (lost) electrons due to its positive (negative) value of $\Delta q$, we found that for both the graphene and GO composites N doping enhances the charge transfer while B doping reduces, and the GO systems exhibit more pronounced charge transfer than the graphene systems.

\begin{figure*}[!th]
\centering
\begin{tabular}{cc}
\includegraphics[clip=true,scale=0.411]{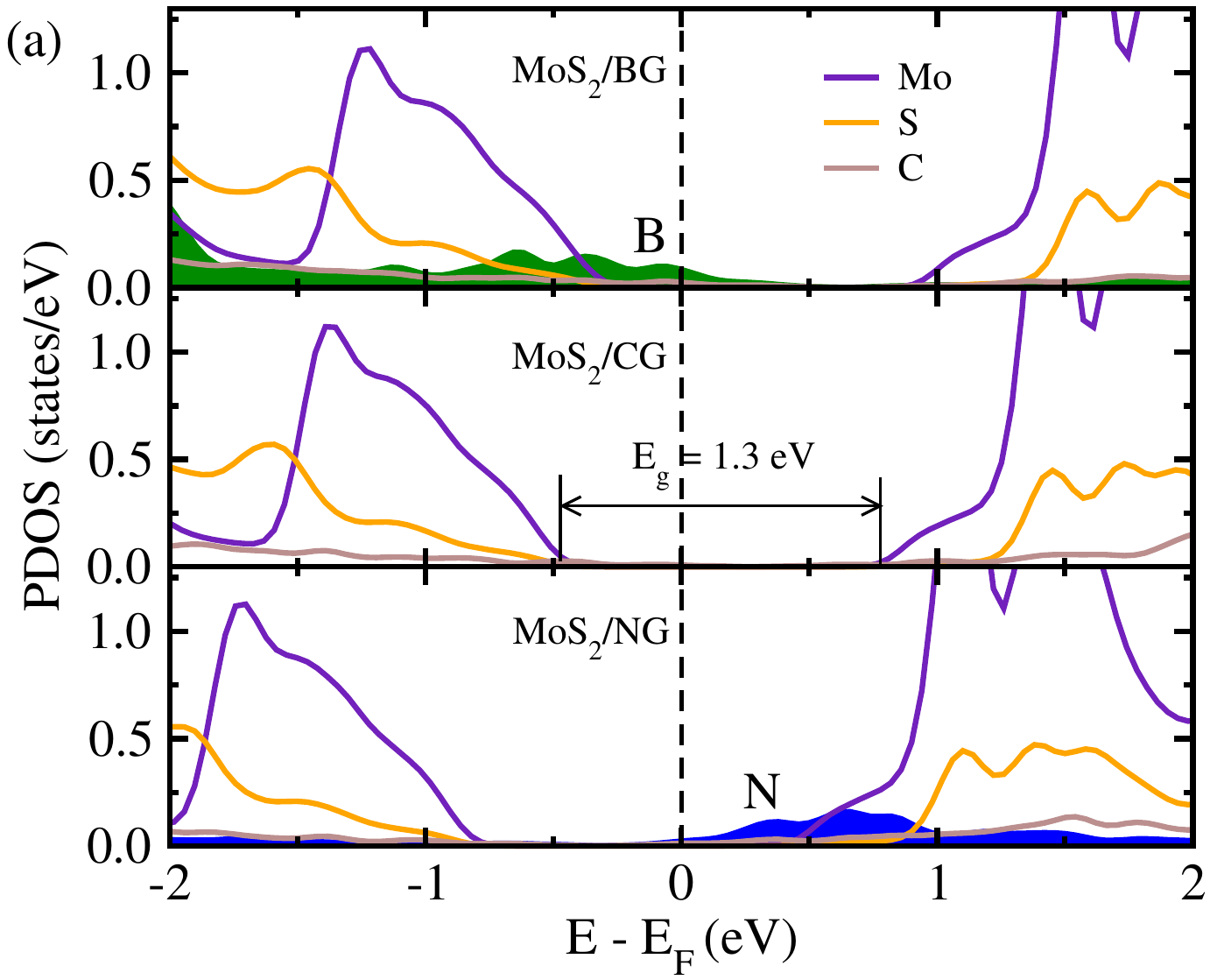} &
\includegraphics[clip=true,scale=0.094]{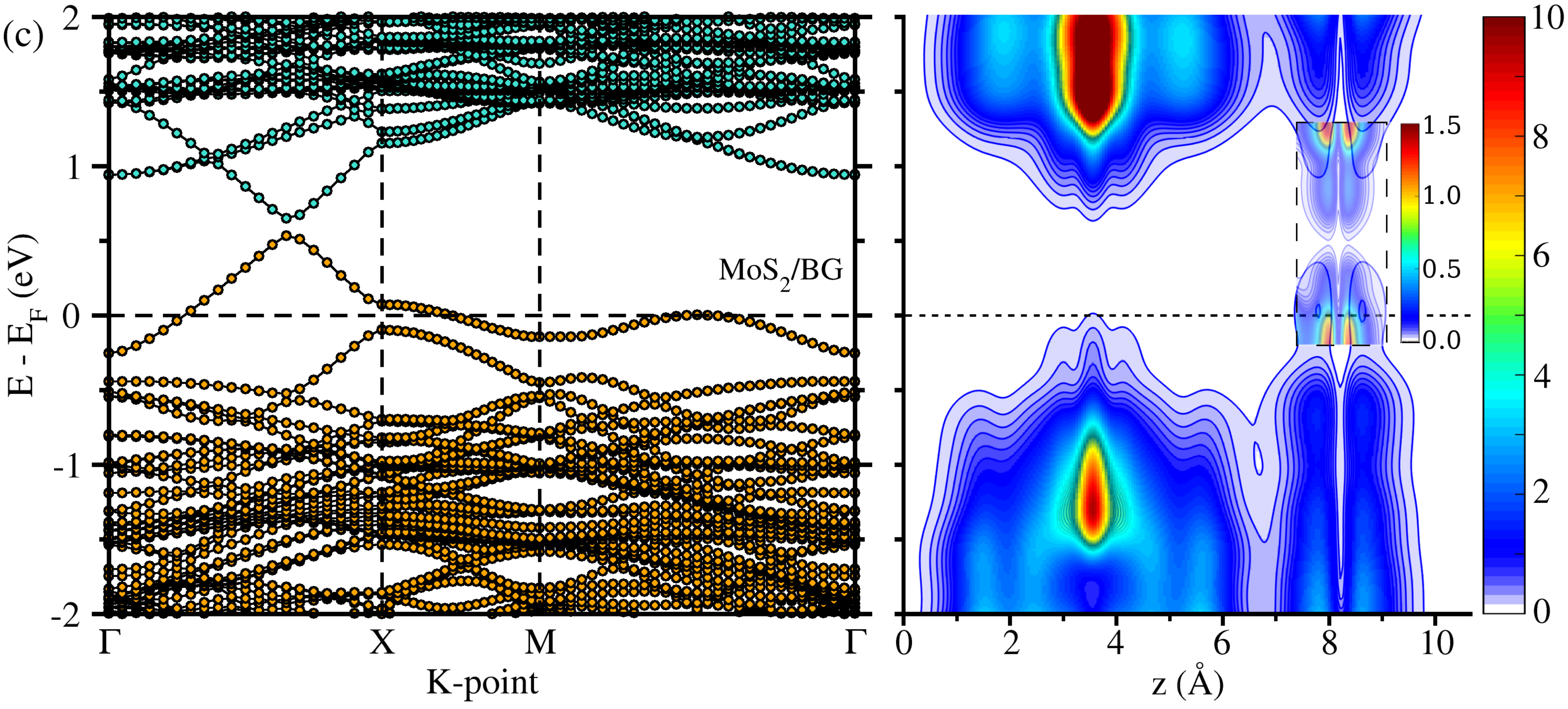} \\
\includegraphics[clip=true,scale=0.411]{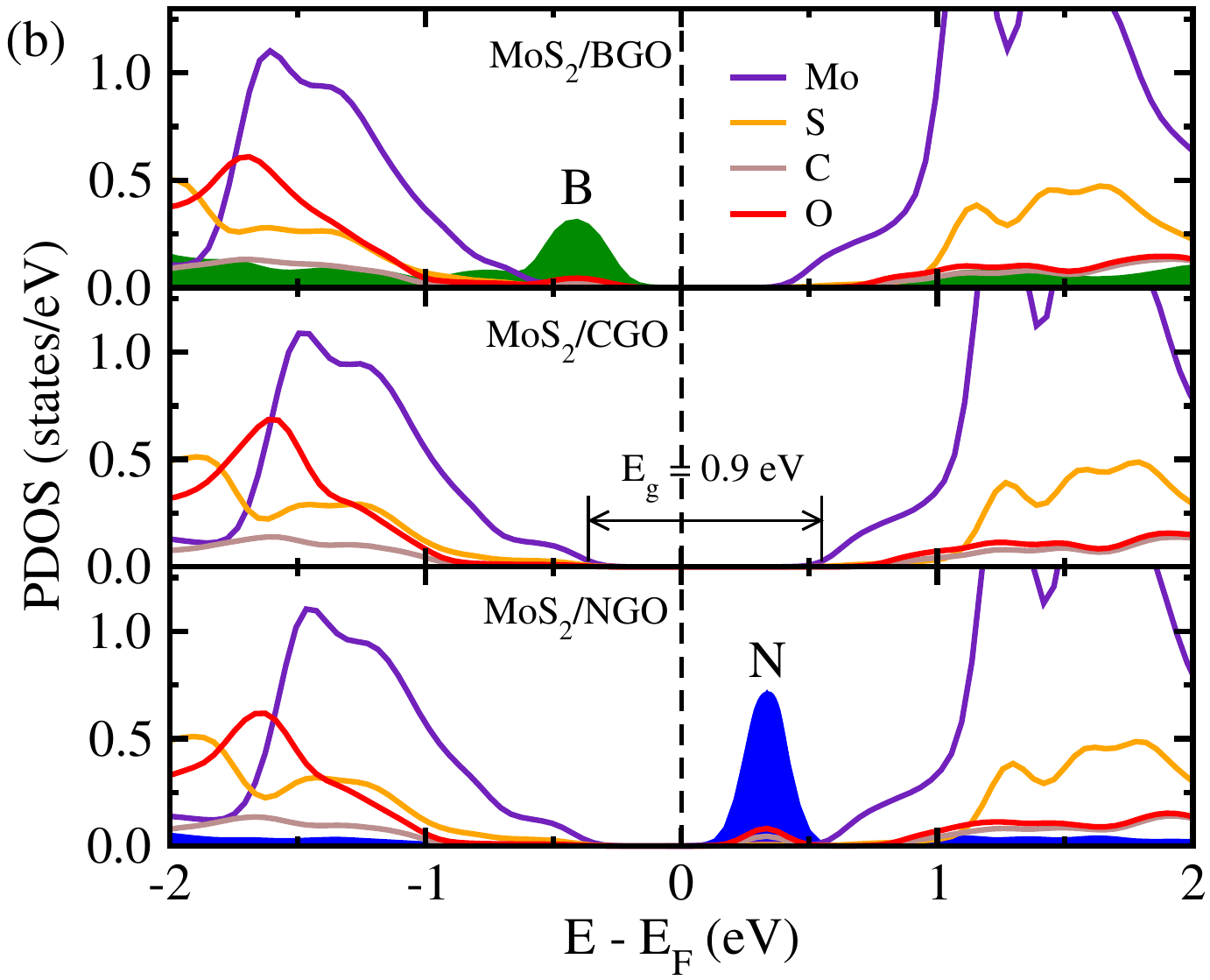} &
\includegraphics[clip=true,scale=0.094]{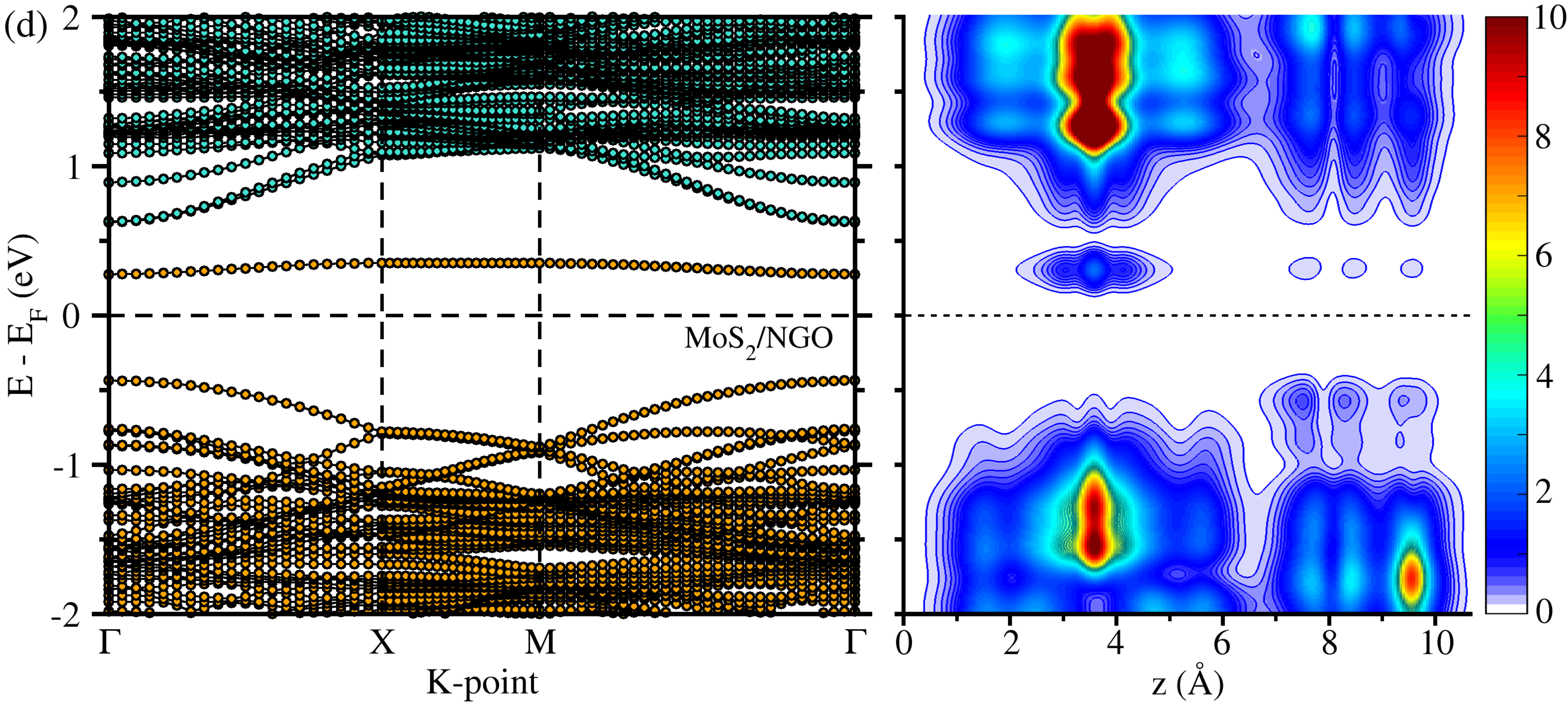}
\end{tabular}
\caption{Partial density of states (PDOS) per atom of \ce{MoS2} interface systems with (a) BG, CG and NG, and (b) BGO, CGO and NGO layers.
The Fermi energy $E_{\text{F}}$ is set zero.
Band structures (left part) and local density of states (LDOS, right panel) of (c) \ce{MoS2}/BG, where the rectangular part indicated by dashed lines was redrawn with detailed scale to clearly show the Dirac cone in the GV layer, and (d) \ce{MoS2}/NGO, selected among the interface systems.}
\label{fig3}
\end{figure*}

The electron redistribution upon the interface formation generates an interface dipole, directing from the \ce{MoS2} layer to the GV layer.
Table~\ref{tab1} presents the interface dipole moments calculated by $\Delta\mu=-\int z\Delta\bar{\rho}(z)dz$, which is divided into the \ce{MoS2} and GV contributions as well.
It turned out that the \ce{MoS2} component of interface dipole with negative value is smaller in magnitude than the GV component with positive value, thereby giving the positive value of total interface dipole for all the interface systems under study.
The dipole moments of 2.25 and 4.20 D for CG and CGO systems increased by N doping to 3.04 and 4.48 D for NG and NGO systems, while decreased by B doping to 1.48 and 3.68 D for BG and BGO systems.

The generated interface dipole, being positive value, results in decrease of work function of the pristine \ce{MoS2} surface.
The work function change can be calculated by $\Delta\phi=e\Delta\mu/\varepsilon_0A$~\cite{Roman13prl} and thus partitioned into two components as listed in Table~\ref{tab1}.
Since the interface dipole is in fact caused by the charge transfer, the work function decrease can be estimated equally by solving a one-dimensional Poisson equation, $d^2\Delta\phi(z)/dz^2=\Delta\bar{\rho}(z)$~\cite{Liu16jpcl,Yu20acsami}.
As illustrated in Figure~\ref{fig2}c, our calculations revealed that the very GV layer brings about the reduction of work function but the \ce{MoS2} substrate increases the work function contrarily.
In accordance with the above findings, the N-doped systems have the most significant reduction, whereas the B-doped systems have the lowest reduction.
In general, N doping into graphene releases non-bonded extra electrons ($n$-type doping), enhancing charge transfer from GV to \ce{MoS2} and thus increasing the interface dipole and work function reduction.
On the other hand, B doping causes deficiency of bonding electrons ($p$-type doping) and thus the electron transfer is reduced.
These indicate that the interfaces can promote the extraction of photo-generated electrons from the \ce{MoS2} (light absorber) and drive them to collect in the GV layer, which can be enhanced by N doping but reduced by B doping.

\subsection{Electronic structure}
To further understand how photo-generated electrons can transfer through the interfaces, we calculated the electronic band structures, partial density of states (PDOS) and local density of states (LDOS), as shown in Figure~\ref{fig3} (see Figures S4 and S5 for all the bilayer composites under study).
The direct band gap of \ce{MoS2} monolayer, which is originated from Mo 4$d$ states and observed at the $\Gamma$ point, was found to decrease from 1.7 eV for the pristine \ce{MoS2} to 1.3 eV for \ce{MoS2}/CG and to 0.9 eV for \ce{MoS2}/CGO.
Such decrease of band gap is attributed to the aforementioned electron transfer from GV to \ce{MoS2}, which is more pronounced for the GO composites.
When compared with the previous calculation using the same DZP basis set~\cite{Ebnonnasir14apl}, our value of 1.3 eV for \ce{MoS2}/CG is smaller by 0.3 eV, possibly due to the slight increase of lateral cell size during matching cell sizes between \ce{MoS2} and GV.
Note that N or B doping barely affects the band gap of \ce{MoS2}, induces up or down shift of the Dirac cone by 0.5 eV for the graphene composites~\cite{Sforzini16prl}, and forms donor or acceptor level.
It is worth noting that the partially occupied bands originated from the GV layer were found to be slightly shifted upward, which can be ascribed to the electron transfer from the GV layer to the \ce{MoS2} layer upon the formation of \ce{MoS2}/GV heterostructure.

\begin{figure}[!b]
\centering
\includegraphics[clip=true,scale=0.62]{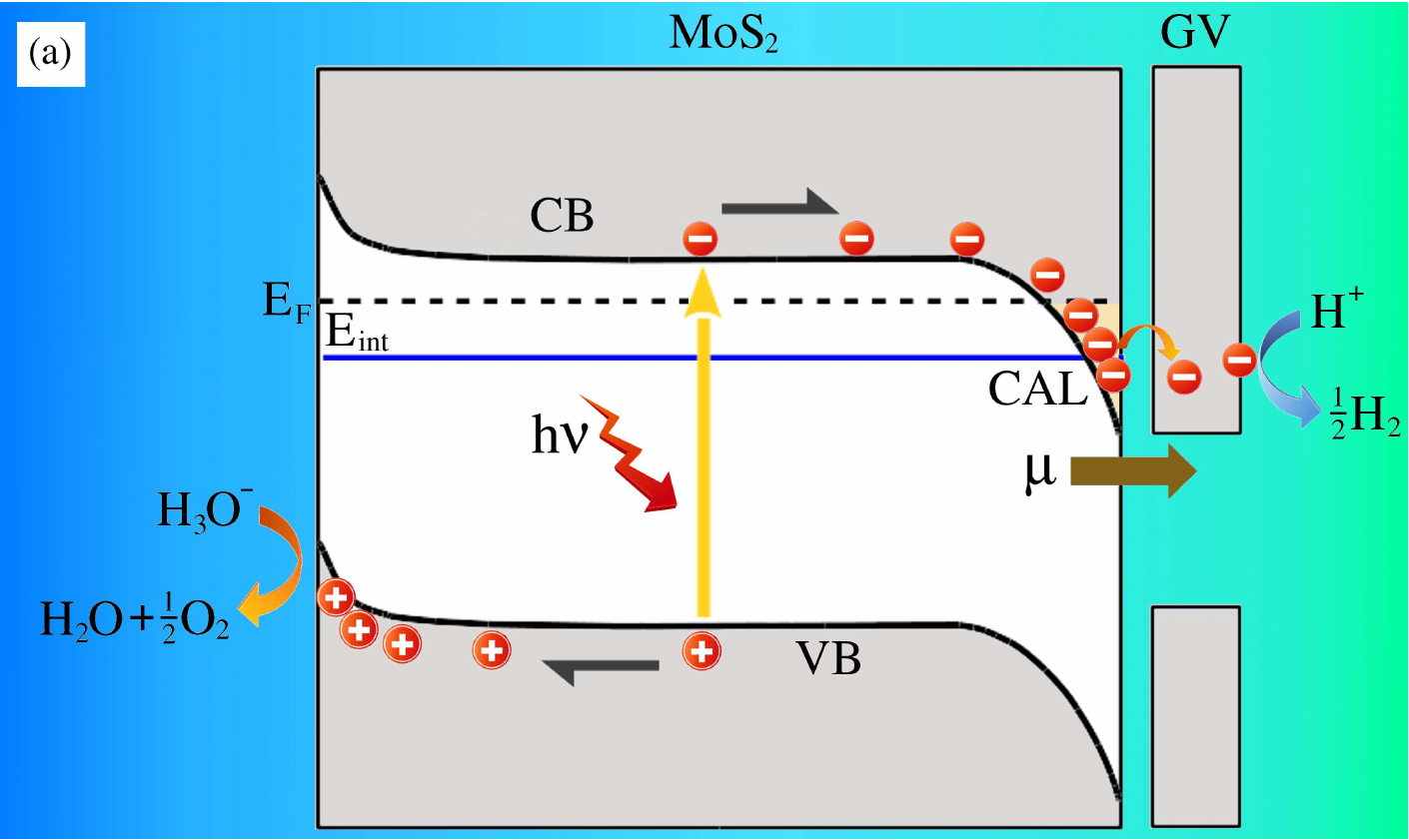} \\ \vspace{5pt}
\includegraphics[clip=true,scale=0.09]{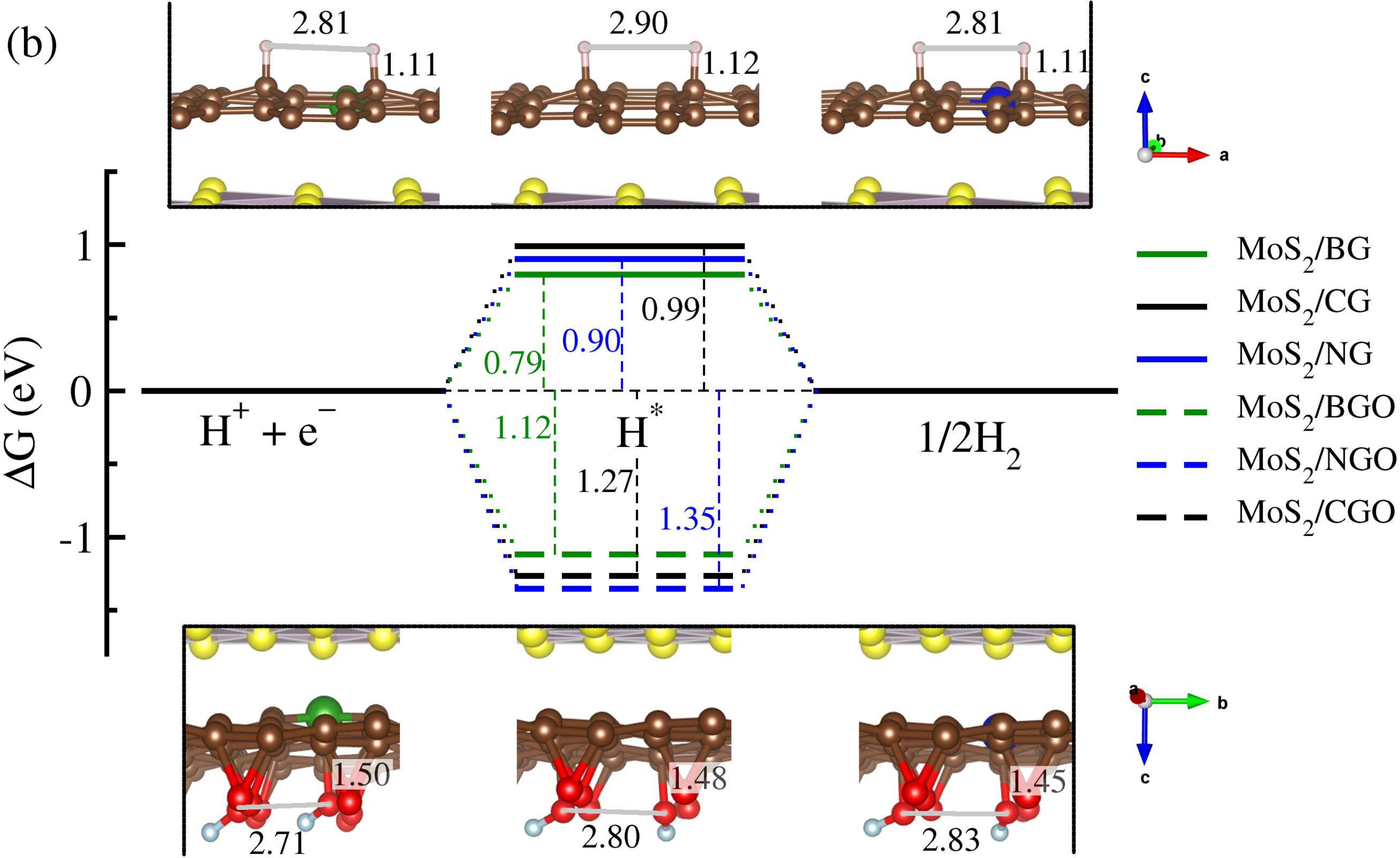}
\caption{(a) Schematic view of photo-generated electron and hole that transfer through conduction band (CB) to GV for hydrogen evolution and valence band (VB) to band edge of \ce{MoS2} for oxygen evolution. Interface dipole ($\mu$) induces downward band bending and thus improves the hopping efficiency of electrons. Interface donor state ($E_{\text{int}}$) enhances formation of charge accumulation layer (CAL) below the Fermi level ($E_{\text{F}}$). (b) Gibbs free energy difference ($\Delta G$) for hydrogen adsorption on the GV side, together with the optimized geometries, where C$-$H and C$-$O bond lengths, and H$-$H and O$-$O interatomic distances are denoted in \AA~unit.}
\label{fig4}
\end{figure}

As \ce{MoS2} captures photons, the valence band electrons are excited to the conduction band, generating the conducting electron and hole, which transfer oppositely along the conduction band and valence band, respectively (see Figure~\ref{fig4}a).
This process is facilitated by the interface dipole and reduced work function, so that the electrons can easily reach the conduction band of GV to induce the hydrogen evolution, \ce{H+} + $e^-$ $\rightarrow \frac{1}{2}$\ce{H2}, while the holes arrive at the valence band edge of \ce{MoS2} to originate the oxygen evolution, \ce{H3O-} + $h^+ \rightarrow \frac{1}{2}$\ce{O2} + \ce{H2O}.
That is, the electron and hole are effectively separated by forming the interface, which is quite beneficial to HER performance.
Notably, for B- and N-doped systems, some new states appear in the band gap of \ce{MoS2} across the interface, i.e., the interface acceptor and donor states, which is clearer in the GO composites (see Figure~\ref{fig3}c-d and Figure S5).
These can be considered as shallow gap states that can provide the recombination center, but the quantum tunneling through these interface states may enhance the photon absorption and offer the carrier tunneling from \ce{MoS2} to GV side~\cite{Qian18apl}.
Accordingly, this quantum effect will increase the transition of photo-generated electrons and enhance the HER performance.
In particular, N doping can form a charge accumulation layer (CAL) below $E_{\text{F}}$, enhancing the downward surface band bending and thus promoting the transition of electrons to the GV layer.

\subsection{Hydrogen atom adsorption}
Finally, based on the finding that photo-generated electrons transfer to the GV side, we studied hydrogen adsorption in the GV side of these heterostructures.
Here, two hydrogen atoms were assumed to be adsorbed on the top of C atoms for the graphene composites, while on the top of O atoms for the GO composites, with different possible sites (see Figure S6).
Figure~\ref{fig4}b shows the optimized geometries of hydrogen-adsorbed interfaces with the lowest total energy.
In the case of \ce{MoS2}/CG composite, the C$-$H bond length and H$-$H distance were found to be 1.12 and 2.90 \AA, protruding the carbon atoms out of the graphene sheet~\cite{Biroju17acs}, and they were slightly decreased to 1.11 \AA~(on the C atom closest to the dopant atom) and 2.81 \AA~by B and N doping.
Meanwhile, for the \ce{MoS2}/CGO composite, two hydroxyl groups with O$-$H bond length of 0.98 \AA~were formed on the same carbon ring (see Figure S7).
With B and N doping, the C$-$O bond length was found to increase and decrease from 1.48 \AA~to 1.50 and 1.45 \AA, while the O$-$O distance to decrease and increase from 2.80 \AA~to 2.71 and 2.83 \AA, respectively.

Gibbs free energy difference ($\Delta G$) for the hydrogen adsorption in \ce{MoS2}/CG, which should be close to zero for the best PEC activity, was calculated to be 0.99 eV, being lower than the previous DFT value of 1.45 eV obtained by adsorbing one H atom~\cite{Biroju17acs} (see Table S1 and S2 for details).
Notably, both B and N doping decrease $\Delta G$ to 0.80 and 0.90 eV, respectively.
For the cases of GO composites, the total energies of hydrogen-adsorbed complexes were found to be lower than the sum of free \ce{H2} and \ce{MoS2}/GV composites, with $\Delta G$ being $-1.12$, $-1.27$ and $-1.35$ eV for the B, C and N variants, respectively.
Therefore, it can be concluded that, although the interface functionality is superior for the GO composite, hydrogen adsorption is preferable on the graphene composite.
Also, N doping enhances more remarkably interface function, but hydrogen adsorption occurs more easily with B doping.
Such contrary effect of B and N doping on interface function and hydrogen evolution deserves much consideration in the design of new functional photocatalysts.

\section{Conclusions}
In summary, we have performed the first-principles calculations of \ce{MoS2} heterostructures with graphene variants such as graphene, graphene oxide and their B- and N-doped derivatives, aiming at clarifying the effect of such doping on interface function and hydrogen evolution.
By using the constructed supercell models for the \ce{MoS2}/GV bilayer heterostructures, we have elucidated their binding characteristics, charge redistribution, interface dipole, work function change, electronic structure and hydrogen adsorption energy, paying attention to comparison between graphene and graphene oxide composites and contrary effect of B and N doping.
It was found that substitution of one carbon atom in graphene and graphene oxide with boron or nitrogen atom substantially modifies the binding nature in the \ce{MoS2}/GV bilayer.
Upon the formation of the bilayer interface, some amount of electronic charge has been found to move from the GV layer to the \ce{MoS2} layer, resulting in the creation of interface dipole and the decrease of work function, being more pronounced in the graphene oxide composites.
Our calculations have shown that N doping enhanced the interface functions with the donor-type interface states but B doping reduces those with the acceptor-type states.
Furthermore, the calculated electronic structures indicated that the interface dipole could facilitate the separation of electrons and holes photogenerated in the \ce{MoS2} layer and caused the transfer of the electrons to the GV layer for hydrogen evolution, which is more promoted by N doping compared to B doping.
We have calculated the hydrogen adsorption energies, revealing that the hydrogen evolution occurred more readily on the B-doped GV side than on the N-doped ones.
We believe this work deserves much consideration in the design of innovative functional photocatalysts for hydrogen evolution.

\section*{Acknowledgments}
The authors thank Y.-H. Kye and J.-S. Kim for their useful comments and valuable help.
Computations have been performed on the HP Blade System C7000 (HP BL460c) managed by Faculty of Materials Science, Kim Il Sung University.

\section*{\label{note}Notes}
The authors declare no competing financial interest.

\bibliographystyle{elsarticle-num-names}
\bibliography{Reference}

\end{document}